\documentstyle[12pt,aaspp4,flushrt]{article} 
\input psfig
\def\simg{\mathrel{\hbox{\rlap{\lower.55ex \hbox {$\sim$}}
                   \kern-.3em \raise.4ex \hbox{$>$}}}}
\def\siml{\mathrel{\hbox{\rlap{\lower.55ex \hbox {$\sim$}}
                   \kern-.3em \raise.4ex \hbox{$<$}}}}
\def\Mesz{M\'esz\'aros~}

\begin{document}
\hfill{\it Subm to ApJL, 7/18/2000}
\title{Delayed X-Ray Afterglows from Obscured Gamma-Ray Bursts\\
  in Star-Forming Regions} 
\author{Peter \Mesz $^{1,2,3}$ and Andrei Gruzinov $^2$}

\affil{$^1$ Institute of Astronomy, Madingley Road, Cambridge CB3 0HA, England, U.K.}
\affil{$^2$ Institute for Advanced Study, School of Natural Sciences, Princeton, 
NJ 08540}
\affil{$^3$ Pennsylvania State University,  525 Davey Lab., University Park, PA 16802~~~~~~~}


\begin{abstract}

For Gamma-Ray Bursts occurring in dense star-forming regions, the X-ray afterglow
behavior minutes to days after the trigger may be dominated by the small-angle scattering 
of the prompt X-ray emission off dust grains. We give a simple illustrative model for 
the X-ray light curves at different X-ray energies, and discuss possible implications.
A bump followed by a steeper decay in soft X-rays is predicted for bursts which are 
heavily obscured in the optical. 

\end{abstract}

\keywords{Gamma-rays: Bursts --- Radiation Mechanisms}

\section{Introduction}

Most of the X-ray fluence of a Gamma-Ray Bursts (GRB) is emitted during the prompt X-ray 
flash, the afterglow emission being smaller by a factor $\sim 10$ (Costa 1999, Piro 2000,
van Paradijs, et al 2000).
For a GRB in a large star forming region, a significant fraction of the prompt X-ray 
emission will be scattered by dust grains. Since the dust grains scatter the X-rays by a 
small angle, time delays of the scattered x-rays will be small (minutes to days, 
depending on the X-ray energy and the grain size). If the X-ray scattering opacity is
substantial, the intermediate timescale, softer part of the X-ray afterglow will be 
dominated 
by the dust scattering, the direct X-ray emission from the blast wave being weaker. This
intermediate X-ray light curve will then generally be steeper than the original 
unscattered afterglow would have been. The optical afterglow will be undetectable, due to 
the high obscuration, but a near-infrared source at the $\mu$Jy level should be present 
over timescales of months.  When a database of X-ray afterglows in different energy 
bands becomes available, through missions such as HETE-2 and Swift, one should be able 
to determine whether some GRBs indeed occur in large dusty regions.

\section{A specific example}

\begin{figure}[htb]
\psfig{figure=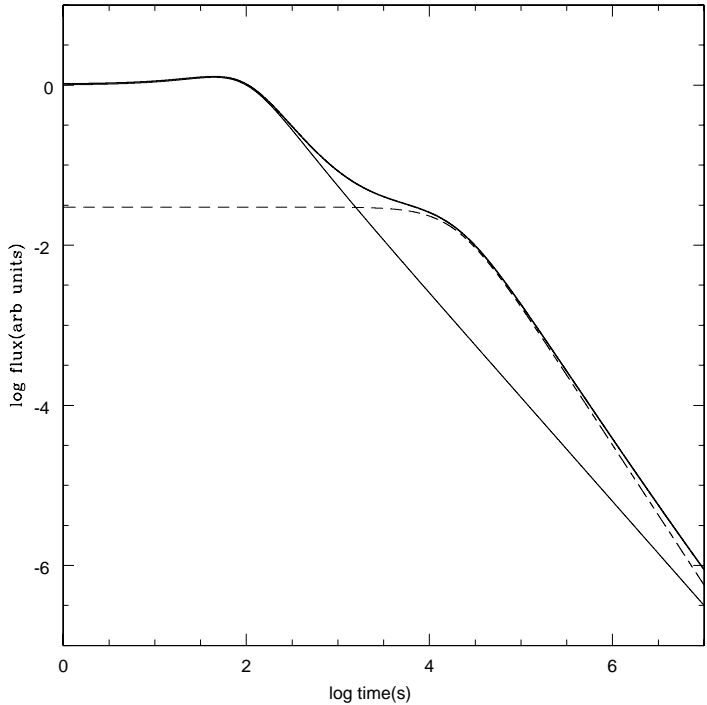,width=4in}
\caption{Dust-scattered X-ray afterglow. Thin line: unscattered 
X-ray flux. Thin dashed line: scattered X-ray flux. Thick line: total flux. The flux 
normalization is arbitrary, while the relative fluxes correspond to the example 
discussed in the text for an energy of 2 keV.}
\label{fig:1}
\end{figure}

To illustrate the phenomenon, we consider as a starting point a typical GRB, whose
unscattered X-ray light curve is parametrized in a simplified manner by two
asymptotic power laws with a peak or break at about 100 s, decaying at late times 
as $t^{-1.3}$,
\begin{equation}
F_0(t)={1+(t/100{\rm s})\over 1+(t/100{\rm s})^{2.3}},
\label{eq:input}
\end{equation}
with an arbitrary normalization depending on the X-ray energy band.
This is represented by the thin line in Fig.\ref{fig:1}.

To make our illustration concrete, we assume that the GRB occurs in a large star forming 
region, of typical radius $R$ about 100pc, where the dust grain populations and optical 
depths are close to what is observed in our Galactic center region. Thus we assume that 
(1) visual extinction is $\sim 10$, (2) X-rays are scattered preferentially by those
dust grains whose size is in the range $a=0.06\mu {\rm m}$, (3) the optical depth to 
dust scattering at the X-ray energy $\epsilon$ is 
\begin{equation}
\tau (\epsilon) = 3 \left( {\epsilon \over 1{\rm keV}}\right) ^{-2}.
\end{equation}
The last two assumptions are adopted from the interpretation of the dust-grain scattering 
observations given by Mitsuda, Takeshima, Kii, \& Kawai (1990). Of course, there is no 
reason to believe that the dust in a distant GRB host galaxy will be similar to the dust 
here. We just use these assumptions as a nominal example.

It follows from assumption (1) that a GRB going off in such a star-forming region will 
have no detectable optical afterglow, but the gamma-rays should be able to penetrate 
through. However the behavior of the X-ray afterglow will be dependent on the
specific energy band being considered.

At X-ray optical depths less than few, dust grains of size $a$ will scatter X-rays of 
energy $\epsilon$ by an angle $\theta \sim 0.2\lambda /a$, where $\lambda$ is the X-ray 
wavelength (Overbeck 1964). We parametrize this as
\begin{equation}
\theta (\epsilon)=4.13\times 10^{-3}\left( {a \over 0.06\mu {\rm m}}\right) ^{-1}\left( {\epsilon \over 1{\rm keV}}\right) ^{-1}.
\end{equation}
The corresponding time lag is $t\sim R\theta ^2/2c$, or
\begin{equation}
t(\epsilon )\sim 8.8\times 10^{4}{\rm s}\left( {a \over 0.06\mu {\rm m}}\right)^{-2}
 \left( {\epsilon \over 1{\rm keV}}\right) ^{-2}\left( {R \over 100{\rm pc}}\right).
\end{equation}
In what follows, we assume that $a=0.06\mu {\rm m}$ and $R=100{\rm pc}$. 

The X-ray telescope (XRT) on the Swift afterglow space mission scheduled to be launched
in 2004 (http://www.astro.psu.edu/xray/swift/xrt/) will be sensitive over an energy 
range $0.2 \leq \epsilon \leq 10$ keV, corresponding to GRB rest-frame energies 
0.5 to 25 keV for a redshift $z=1.5$. Let us consider the dust contribution at several 
GRB-frame X-ray energies. 

At 10 keV, the X-ray optical depth given by equation (2) is much less than ubity, so 
the X-ray afterglow is not affected by the dust scattering. The light curve will then 
just be the usual unscattered time dependent X-ray afterglow, which in this example is
parametrized through equation (1).

At 3 keV, the optical depth is $\tau \sim 0.3$. The time lag, equation (4), is 
$t\sim 10^4$s. Since the total fluence is $f\sim \int dt F_0\sim 500$s, the scattered 
flux is $F_s\sim \tau f/t\sim 0.01$. The unscattered flux at $10^4$s is given by 
equation (1), $F_0\sim 3\times 10^{-3}$. Thus scattering has only a minor influence on 
the afterglow.

At 2 keV, the optical depth is $\tau \sim 1$. The time lag is $t\sim 2\times 10^4$s. 
The scattered flux is $F_s\sim \tau f/t\sim 0.03$. The unscattered flux at $2\times 
10^4$s is $F_0\sim 10^{-3}$. In the time interval from hours to weeks, the dust 
scattering dominates the afterglow, and, as shown in Fig.\ref{fig:1} , the afterglow 
is approximately a power law $F\propto t^{-1.75}$. This is because dust grains of 
radius $a<0.06\mu {\rm m}$ will scatter the prompt emission with longer time lags, 
$t\propto a^{-2}$, and with smaller optical depths $\tau$. To calculate $\tau$, we 
take a standard dust grain size distribution where the number of grains of size of 
order $a$ is $\propto a^{-2.5}$ (Mathis, Rumpl, \& Nordsieck 1977). For a scattering 
cross section $\propto a^4$ (Overbeck 1964), the optical depth is $\tau \propto 
a^{1.5}\propto t^{-0.75}$, so the flux $F\propto t^{-1.75}$.

At 1 keV, the optical depth is $\sim 3$, and the dust grain scattering dominates the 
afterglow at $t>10$min. Multiple scatterings will be important, the net deflection
angle being, in the small-angle deflection regime, ${\bar \theta} \propto N_{sc}^{1/2} 
\theta(\epsilon)\propto \tau^{1/2}(\epsilon) \theta(\epsilon)\propto \epsilon^{-2}$, 
so the time delay is $t(\epsilon)\sim {\bar\theta}^2(R/2c) \propto \epsilon^{-4}$. 
This decreases at early times the ratio of the scattered to the unscattered flux, 
compared to the values in the single scattering regime at 2 keV, and it increases 
the time after which the scattered flux becomes dominant, the late time decay having 
the same time exponent of -1.75.

\section{Discussion}

The specific example that we have described shows that highly obscured star-forming 
regions should lead to specific signatures in the X-ray light curve of gamma-ray
bursts occurring in them. This consists of a secondary flattening or bump in the 
light curve at X-ray energies $\epsilon \sim 2-3$ keV. Depending on the size and dust 
column density of the region, this bump occurs hours to days after the initial 
``canonical" X-ray decay has been going on, followed by a steeper $F_x\propto t^{-1.75}$ 
decay. At lower energies (e.g. in the 0.2-0.5 keV range) the X-ray bump 
in the light curve should appear at increasingly later times and contribute a 
decreasing fraction of the total energy. 

The presence of this X-ray signature is expected to be associated with bursts which do 
not produce a detectable transient at optical wavelengths (OT). There are at
least three bursts so far for which an X-ray decay index is known and an OT was not
detected: GRB 970204, 970828 and 991214. The X-ray decay  indices of these were
-1.4, -1.56 and -1.00, based on GCN notices (e.g. Greiner, 2000). The third is
clearly a canonical decay, and should be the unscattered component, but the other
two could be the scattered component, considering the uncertainties in the grain
size distribution and cross sections that determine the decay rate. There is a larger
number of bursts for which an X-ray afterglow was reported, while an OT was not
found. However, since optical observations first require an X-ray position,
which so far has been possible only after hours of delay and with arc-minutes accuracy,
there is currently a possible bias against finding OTs.  Stronger constraints will 
have to await the faster coordinate alerts and smaller X-ray error circles expected 
from dedicated GRB afterglow missions such as HETE-2 and Swift.

An infrared source is expected to be associated with such obscured, X-ray peculiar 
GRBs, since the dust will re-radiate the UV and soft X-ray radiation of the absorbed 
source.  Thermal reemission and scattering outside the sublimation radius is expected 
to cause delayed IR emission even when the optical transient is only partially absorbed 
(Waxman \& Draine 2000, Esin \& Blandford 2000), and the same is expected here, except 
for the optical transient being compeletely obscured. As a numerical example, for an 
isotropic equivalent total burst energy $E\sim 10^{53}$ erg at a redshift $z\sim 1$ the 
normalization of the X-ray flux for the burst of Figure 1 would be $F_x\sim 10^{-9}$ 
erg cm$^{-2}$ s$^{-1}$ keV$^{-1}$ for $t\siml 100$ s, in the usual range of X-ray 
afterglow fluxes detected by Beppo-SAX (Costa, et al, 1999). The dust reradiation occurs 
beyond the sublimation radius $R_s\sim 10~L_{49}^{1/2}$ pc at wavelengths $\lambda\simg 
2(1+z)\mu$m, where $10^{49}L_{49}$ erg/s is the early UV component of burst afterglow 
(Waxman \& Draine 2000). The time delay associated with the reradiated flux is $t_{IR} 
\sim (R_s/2c) \theta_j^2$ where $\theta_j=10^{-1}\theta_{-1}$ is a typical collimation 
half-angle of the burst radiation. At $z\sim 1$ the corresponding infrared flux at 2.2 
$\mu$m would be $F_{2.2\mu\hbox{m}} \sim L_{49} \theta_j^2 /[4\pi D_L^2 (R_s/2c) 
\theta_j^2 \nu ]\sim 0.3 L_{49}^{1/2}$ $\mu$Jy, independent of $\theta_j$, or 
$m_K\sim 23.3$ compared to Vega, approximately constant for a time $t_{IR} \sim 
5\times 10^6\theta_{-1}^{2}L_{49}^{1/2}$ s. The IR flux of the host galaxy at that 
redshift could exceed this value, but 8-m class telescopes in good seeing conditions or 
with adaptive optics would resolve the galaxy, facilitating detection of the point-like 
IR afterglow. If the X-ray to IR fluxes can be calibrated for a sample of sources at 
$z\sim 1$, such $\gamma$-ray detected GRBs with anomalous X-ray afterglow behavior and 
no OT may be used as tracers of massive stellar collapses. It may thus be possible to 
detect star-forming regions out to redshifts larger than those detectable with optical
or infrared techniques, since typical GRB $\gamma$-ray and X-ray fluxes can in 
principle be measured out to $z\sim 10-15$.

\acknowledgements{We thank Andrew Blain, Bruce Draine, Masataka Fukugita, David Hogg,
Kevin Hurley, Richard McMahon, Martin Rees and Eli Waxman for useful discussions. 
PM was supported by NASA NAG5-9192, the Guggenheim Foundation, the Sackler Foundation 
and the Institute for Advanced Study. AG was supported by the W. M. Keck Foundation 
and NSF PHY-9513835. } 
 
\eject

\end{document}